\documentclass[pre,showkeys,preprintnumbers,amsmath,amssymb,superscriptaddress,twocolumn]{revtex4}

\usepackage{graphicx}
\usepackage{amsfonts}
\usepackage{url}
\usepackage{yfonts}

\usepackage{epstopdf}
\usepackage{nameref,hyperref}
\usepackage{comment}
\usepackage{amsmath} 
\usepackage{lastpage,fancyhdr,graphicx}
\usepackage{color}

\newcommand{\blue}[1]{\textcolor{blue}{#1}}

\begin{document}
\title{Continuous limits of Heterogeneous Continuous Time Random Walk}


\author{Liubov Tupikina}
\email{liubov.tupikina@cri-paris.org}
\affiliation{University Paris Descartes,
	 Centre of Research and Interdisciplinarity,  INSERM,  France}
\affiliation{Nokia Bell labs, France}

%\author{Denis~S.~Grebenkov}
%\email{denis.grebenkov@polytechnique.edu}
%\affiliation{	Laboratoire de Physique de la Mati\`{e}re Condens\'{e}e (UMR 7643), \\ 	CNRS -- Ecole Polytechnique, 91128 Palaiseau, France}

\date{\today}

%\tableofcontents

\begin{abstract}
	Continuous time random Walk model has been versatile analytical formalism for studying and modeling diffusion processes in heterogeneous structures, such as disordered or porous media.
	We are studying the continuous limits of Heterogeneous Continuous Time Random Walk model, when a random walk is making jumps on a graph within different time-length. 
	We apply the concept of a generalized master equation to study heterogeneous continuous-time random walks on networks.
	Depending on the interpretations of the waiting time distributions the generalized master equation gives different forms of continuous equations.
\end{abstract}

\maketitle

\section{Introduction}

The continuous time random walk (CTRW) model has been widely used 
for modelling dynamics inside porous media \cite{Scher1973}.
However homogeneous  random walk models do not exhaust the whole variety of dynamical phenomena \cite{Sahimi2012,Berkowitz2000}. 
The effects of heterogeneities on stochastic dynamics have been investigated in random trap and barrier models \cite{Thiel2016, Sokolov2010}. Recently there have been several studies, which describe models of random walks in heterogeneous medium.  
In particular, in Heterogeneous Continuous Time Random Walk (HCTRW) model on a network, introduced in
\cite{GrebTupikina}, %discrete structure,
 a random walk is moving from one site $x$ to $x'$ with the waiting time distribution $\psi_{xx'}(t)$ depending on the link.  
HCTRW model allows one to introduce the local heterogeneity encoded through the distributions of travel times \cite{TuGrebenk}, therefore it is of the general interest to develop a continuous equivalent 
of HCTRW model and explore its potential for description of diffusion models.

%on on  lattices or irregular graphs
The fractional diffusion equations were recognized as a useful tool for the description of anomalous diffusion in disordered medium  \cite{Chechkin2006}.
In some  cases corresponding fractional differential equations have been investigated \cite{Hughes, Nigris}. For homogeneous CTRW, when all waiting time distributions are  fixed, %$\psi(t)$ 
the corresponding continuous dynamics is described by fractional diffusion equation with a Riemann-Liouville fractional derivative  
\cite{Barkai, Schumer}. %Serov's paper
However for cases when macroscopic behaviour is heterogeneous and obeys some local laws, continuous limits description
is a challenging mathematical problem  \cite{Klafter1980}.

In this article we are describing random processes with incorporated heterogeneities, we are studying corresponding continuous limits of a discrete model. 
For this we first derive discrete equations with  continuous limits for
HCTRW framework \cite{GrebTupikina}.
In particular, we derive the continuous time limit for the HCTRW on large graphs. Our results for HCTRW generalize previously known equations  for lattices and off-lattice walks \cite{Sokolov2011}. %\cite{Mohar1989}.
In Section \ref{sec_deriv} we first derive the most general form of continuous limits for the Heterogeneous Continuous Time Random Walk (HCTRW) model dynamics. Then in Section \ref{sec_cases}
we introduce and derive various types (or interpretations) of continuous limits of HCTRW. 
As an outlook, in Section \ref{sec_disc} we discuss possible connections between interpretations of HCTRW model and various formalisms for diffusion equations. Finally in Section \ref{sec_concl} we conclude and make the overview about future steps. % answering some of open questions for HCTRW framework.

\section{Continuous limits for HCTRW dynamics}
\label{sec_deriv} 
\subsection{HCTRW model}

%\blue{Change $x_i \rightarrow \bar{x}$.}

\begin{figure}
	\includegraphics[width=1.7in]{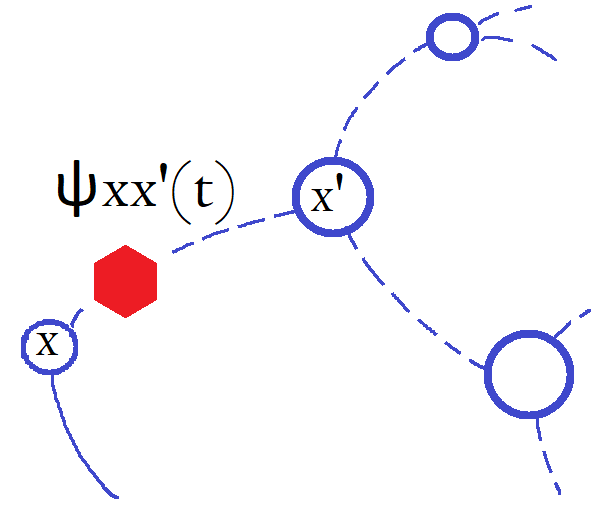}\centering
	\caption{Illustration of the HCTRW model. A random walk jumps from site $x$ to $x'$ during time, which is a random variable drawn from the probability density  $\psi_{xx'}(t)$. }	
	\label{fig_reg}
\end{figure}

In \cite{GrebTupikina} we introduced the Heterogeneous Continuous Time Random Walk (HCTRW) model  a random walk moves on a graph (or a network) $G$ in continuous time jumping from one node
to another set by a transition (stochastic) matrix $Q$ whose element
$Q_{xx'}$ is the probability of jumping from the node $x$ to $x'$ via link $e_{xx'}$ and the travel time needed to move along this link is a random variable drawn from the probability density $\psi_{xx'}(t)$.
%The HCTRW model on a graph with $N$ nodes is set by a transition matrix $Q$ of size $N\times N$, the heterogeneous distributions of travel times, denoted by $\psi_{\bar{x}x'}(t)$ for each edge $e_{\bar{x}x'}$.
The coupling between spatial and temporal properties of random walk dynamics is set 
by the elements of a generalized transition matrix $Q(t)$: 
$Q_{\bar{x}x'}(t) = \psi_{\bar{x}x'}(t)Q_{\bar{x}x'}$. 
A snapshot of the HCTRW model is  schematically shown on Fig.~\ref{fig_reg}.

\subsection{Derivation of continuous limits for HCTRW dynamics}
There are different ways to derive continuous limits from a random walk microscopic dynamics
 \cite{Landman1977, Klafter1980, Sokolov2007}. 
%(continuous limit for HCTRW on a graph)
First we start with the heuristic derivation of the generalized master equation (GME). GME is the integro-differential
equation that describes the evolution of a system in
time \cite{Sokolov2011}, 
which is often used for derivation of Fokker-Planck diffusion equations. %We derive the generalized Fokker-Planck equation for the propagator $P_{x_0x}(t)$ for continuous space 
The generalized master equation is based on two balance conditions:
(i) the local balance between the gain and loss fluxes at each site; %which is independent from an initial position $x_0$
(ii) the balance of transitions
between any two sites, representing a particle conservation during jumps, e.g. the continuity property. These two conditions guarantee the probability conservation. %we will use the balance equation for fluxes in order to describe the density evolution of  $P_{x_0x}(t)$ \cite{Sokolov2011}.
We denote the probability to find a particle at time $t$ at site $\bar{x}$ and initially located at site $x_0$ by $P_{x_0\bar{x}}(t)$. 
Using the former condition (i) we represent the balance  equation for the HCTRW as the standard balance equation between loss and gain fluxes  $j^{\pm}_{\bar{x}}(t)$ in a site $\bar{x}$:
\begin{align}
\frac{d P_{x_0\bar{x}}(t)}{d t} = j^+_{\bar{x}}(t) - j^-_{\bar{x}}(t).
\label{eq_bal}
\end{align}
The probability of a particle leaving   $\bar{x}$ between $t$ and $t+dt$ is $j^-_{\bar{x}}(t)dt$. % (the probability of leaving is $j^-_x(t)dt$). 
There are two possible scenarious: a particle leaving site $\bar{x}$ between $t$ and $t+dt$ 
either stays at site $\bar{x}$ from $t=0$ and in this case $\bar{x}=x_0$, or 
a particle arrives to  site $\bar{x}$ at the later time $t'$ such that $0<t'<t$  
%Then the total probability of leaving between $t$ and $t+dt$ is expressed as $\sum_{x'}q_{xx'}\psi_{xx'}(t) P_{x_0x}(0)dt  + (\sum_{x'}\int_{0}^{t}q_{xx'} \psi_{xx'}(t-t')j^+_x(t')dt')dt, $. 
for a discrete case. 
Therefore the loss flux is expressed as %, using Markovian property, 
\begin{align}
j^-_{\bar{x}}(t) = \sum_{x'}Q_{\bar{x}x'}\psi_{\bar{x}x'}(t) P_{x_0\bar{x}}(0) + \\ \nonumber  \int_{0}^{t}\sum_{x'}Q_{\bar{x}x'} \psi_{\bar{x}x'}(t-t')j^+_{\bar{x}}(t')dt',
\label{eq_flux_neg}
\end{align} 
where 
$P_{x_0\bar{x}}(0)$ is the initial probability distribution.  
In both sums we consider nodes $x'$ adjacent to node $\bar{x}$, 
since the contribution to the total flux from each node $x'$ is given by the term $Q_{\bar{x}x'}\psi_{xx'}(t-t')j^+_{\bar{x}}(t')$.
For convenience we put $\mathcal{Q}_{\bar{x}}(t) =\sum_{x'}Q_{\bar{x}x'}\psi_{\bar{x}x'}(t)$. % which is related to $\Psi_{\bar{x}}(t)$ from \cite{GrebTupikina}. 
Then the expression for the loss flux becomes: %Eq.~(\ref{eq_flux_neg}) 
\begin{align}
j^-_{\bar{x}}(t) = \mathcal{Q}_{\bar{x}}(t) P_{x_0\bar{x}}(0) + \int_{0}^{t}\mathcal{Q}_{\bar{x}}(t-t')j^+_{\bar{x}}(t')dt'.
%\label{eq_flux}
\end{align}
Then the expression for the flux in Laplace domain is:% to Eq.~(\ref{eq_flux_psi})
%\begin{equation}
%j^-_{\bar{x}}(t) = \mathcal{Q}_{\bar{x}}(t){P}_{x_0\bar{x}}(0) + \int_{0}^{t} \mathcal{Q}_{\bar{x}}(t-t')\big(\frac{d {P}_{x_0\bar{x}}(t')}{d t}  + j^-_{\bar{x}}(t')\big)dt',
%\label{eq_flux_psi}
%\end{equation}
\begin{align}
\tilde{j}^-_{\bar{x}}(s) = \tilde{\mathcal{Q}}_{\bar{x}}(s){P}_{x_0\bar{x}}(0) + 
\\ \nonumber \tilde{\mathcal{Q}}_{\bar{x}}(s)(s\tilde{P}_{x_0\bar{x}}(s) -{P}_{x_0\bar{x}}(0) + \tilde{j}^-_{\bar{x}}(s)).
\end{align}
From here on the tilde denotes Laplace-transform of the corresponding function. 
The expression for $\tilde{j}^-_{\bar{x}}(s)$ allows us to express $j^-_{\bar{x}}(t)$ through  ${P}_{x_0\bar{x}}(t)$ and $\mathcal{Q}_{\bar{x}}(t)$ in Laplace domain: 
\begin{align}
\tilde{j}^-_{\bar{x}}(s) =s\frac{\tilde{\mathcal{Q}}_{\bar{x}}(s)}{1-\tilde{\mathcal{Q}}_{\bar{x}}(s)} \tilde{P}_{x_0\bar{x}}(s) \equiv \tilde{\Phi}_{\bar{x}}(s)\tilde{P}_{x_0\bar{x}}(s),
\label{eq_j_s}
\end{align}
where the memory kernel  $\tilde{\Phi}_{\bar{x}}(s)$ is expressed through  $\tilde{\mathcal{Q}}_{\bar{x}}(s)$, %sets the evolution of $\tilde{P}_{x_0\bar{x}}(s)$ in Laplace domain. %$\tilde{\mathcal{Q}}_{\bar{x}}(s)$.
which
intrinsically depends on travel time distributions $\psi_{\bar{x}x'}(t)$ between ${\bar{x}}$ and neighboring sites $x'$. The loss flux also intrinsically depends on a local structure of a network and temporal heterogeneities defined by the generalized transition matrix $Q(t)$ \cite{GrebTupikina}. %and does not depend on the initial condition $P_{x_0\bar{x}}(0)$?
For convenience we introduce function $M_x(t)$, which in Laplace domain gives:
\begin{align}
\tilde{M}_{\bar{x}}(s) = \frac{\tilde{\mathcal{Q}}_{\bar{x}}(s)}{1-\tilde{\mathcal{Q}}_{\bar{x}}(s)},
\label{eq_m_x}
\end{align} 
which we are using furthermore to dissect various generalized  master equations. 
%Then the inverse Laplace transform of  Eq.~(\ref{eq_j_s}) gives us the expression for the flux 
%\begin{align}
%j^-_{\bar{x}}(t) = \int_{0}^{t}
%\hat{\Phi}_{\bar{x}}(t-t') P_{x_0\bar{x}}(t')dt %=\mathcal{L}^{-1} (sM_x(s)\tilde{P}_{x_0x}(s))
%= %% % NO \nonumber!!
%\frac{d }{dt}\int_{0}^{t} ( M_{\bar{x}}(t-t')) P_{x_0\bar{x}}(t')dt',
%\label{eq_M_x}
%\end{align}
%where the integro-differential operator $\hat{\Phi}_{\bar{x}}$  is expressed through the kernel $M_{\bar{x}}(t)$  is the inverse Laplace transform  of $M_{\bar{x}}(s)$. % The memory kernel  $M_{\bar{x}}(t)$ is the integral of  $\hat{\Phi}_{\bar{x}}$.
%We check that 
%	$\mathcal{L}(\frac{d }{dt}\int_{0}^{t} ( M_{\bar{x}}(t-t')) P_{x_0{\bar{x}}}(t')dt'$ gives $sM_{\bar{x}}(s)\tilde{P}_{x_0\bar{x}}(s)$.

The condition  (ii) on the probability conservation of fluxes between sites  can be written in the form, where  the gain flux received at ${\bar{x}}$ consists from the loss fluxes from adjacent sites weighted by the transition probabilities: 
\begin{align}
j^+_{\bar{x}}(t) = \sum_{x'}Q_{{\bar{x}}x'} j^-_{x'}(t).
\label{eq_flux_cons}
\end{align}
Writing
Eq.~(\ref{eq_flux_cons}) in Laplace domain together with Eq.~(\ref{eq_bal}) gives us the final expression for GME in Laplace domain  
\begin{align}
s\tilde{P}_{x_0\bar{x}}(s) -P_{x_0\bar{x}}(0) = \\ \nonumber \sum_{x'}Q_{\bar{x}x'}\Big(  s\tilde{M}_{x'}(s) \tilde{P}_{x_0x'}(s) \Big) - 
s\tilde{M}_{\bar{x}}(s) \tilde{P}_{x_0\bar{x}}(s).
\label{eq_dif_lapl}
\end{align}
%where the term for the initial condition on the right hand side cancels due to \blue{ stochastic matrix property.}
While in the real time domain we can write:
\begin{align}
\dot{P}_{x_0\bar{x}}(t) = \sum_{x'}Q_{\bar{x}x'}\Big(\frac{d}{dt} \int_{0}^t  M_{x'}(t-t') P_{x_0x'}(t') dt'\Big) - 
\\ \nonumber
\frac{d}{dt} \int_{0}^t  M_{\bar{x}}(t-t') P_{x_0\bar{x}}(t')dt'.
\label{eq_fin_GME}
\end{align}

For simplicity, we first consider GME for the HCTRW on a regular graph and put $Q_{\bar{x}x'}=1/2$, if other value is not stated explicitly. For any $k$-regular graph with fixed degree $k$ for each node  we set $Q_{\bar{x}x'} = \frac{1}{k}$.
Then from Eq.~(\ref{eq_fin_GME}) we get the GME
in a simpler form %\green{  on an interval }
\begin{align}
\dot{P}_{x_0\bar{x}}(t) =  \frac{d}{dt} \int_{0}^t \big( \frac{1}{2}\sum_{x'=\bar{x}\pm 1} M_{x'}(t-t') P_{x_0x'}(t') - \\ \nonumber
-M_{\bar{x}}(t-t')P_{x_0\bar{x}}(t')\big)dt'.
\label{eq_GMEgen}
\end{align}

Further, in Section \ref{sec_cases} we derive  the continuous limits for various setups (or interpretations) of the HCTRW  model.
%In the section after we also discuss the diffusion equation conventions.

\section{Interpretations of HCTRW}
\label{sec_cases}
Here we  
introduce various setups (or interpretations) of the HCTRW model, 
 depending on the definition of the travel time distributions $\psi_{\bar{x}x'}(t)$, Fig.~\ref{fig_reg}.
We consider two phases of a jump of the HCTRW model:  (1)
when a random walk decides where to go with the fixed transition probability; (2) when a random walk decides how long does it take. 
All in all, we distinguish three different HCTRW model interpretations: 
\\
$\bullet$
\emph{ The first (I) continuous HCTRW interpretation} %\blue{(change it to \textbf{I})}: 
a random walk stays in node $\bar{x}$ during the travel time, which in this case is denoted as  $\psi_{\bar{x}}(t)$, 
and then instantaneously jumps to $x'$. 
\\
$\bullet$ \emph{ The second (II) continuous HCTRW interpretation}:  %(\blue{Change the order of interpretations: second $\rightarrow$ third})
a random walk arrives to a node $\bar{x}$ and stays on an edge between $\bar{x}$ and $x'$
during time drawn from the probability density function $\psi_{\bar{x}x'}(t)$.  %(i.e. at $(\bar{x}+x')/2$)
\\
$\bullet$ 
\emph{ The third (III) continuous HCTRW interpretation}: a random walk arrives to $x'$   from $\bar{x}$  and waits at $x'$ during time driven with the probability density function, in this case denoted as $\psi_{x'}(t)$. 

The interpretations defined above also correspond to different cases of, so-called, up- and down-times of links activations of stochastic temporal networks \cite{Petit2019}.
Here we attempt to rigorously define the continuous limits of the HCTRW model
%Further we make calculations starting from  Eq.~(\ref{eq_dif_lapl}) in  Laplace domain.
and the term "interpretation" (or "continuous interpretation") should be understood as a microscopic level of interpretation.
The Langevin equation can be used to describe %aggregates 
the separate characteristics of a physical system, necessary for homogeneous and heterogeneous systems.
We note that in the context of the Langevin equation the term "interpretation" was used in order to describe the inverse procedure to the aggregation
\cite{Sokolov2010}.

\subsection{First HCTRW interpretation }
\label{subsec_1interpr}
We start with  \emph{the first HCTRW interpretation} on one dimensional lattice, 
when the travel time distribution depends only on one node $\bar{x}$ (from where the HCTRW comes from) and not on $x'$:   $\psi_{\bar{x}x'}(t)= \psi_{\bar{x}}(t)$.    
Then Eq.~(\ref{eq_m_x}) becomes 
$$
\tilde{M}_{\bar{x}}(s) = \frac{\mathcal{Q}_{\bar{x}}(s)}{1-\mathcal{Q}_{\bar{x}}(s)} = \frac{\tilde{\psi}_{\bar{x}}(s)\sum_{x'}Q_{\bar{x}x'}}{1-\tilde{\psi}_{\bar{x}}(s)\sum_{x'}Q_{\bar{x}x'}}
=\frac{\tilde{\psi}_{\bar{x}}(s)}{1-\tilde{\psi}_{\bar{x}}(s)}.
$$
%valid for all stochastic matrices $Q$. %where we used the property of stochastic matrix $Q$. 

For the first HCTRW interpretation
we consider various cases of dependence of $\psi_{\bar{x}}(t)$ and $\bar{x}$: 
(o) $\psi_{\bar{x}}(t) =\psi(t)$ for all nodes $ \bar{x}$; 
(i) all travel time distributions are exponential with parameter depending on the node $\bar{x}$; (ii) all travel time distributions have finite moments; 
(iii) at least one moment of travel time distribution is infinite.

(o) The simplest case is when all travel time distributions are the same
$\psi(t)$ for all nodes $ \bar{x}$. Then  Eq.~(\ref{eq_fin_GME}) is transformed to the form: % finite difference equation 
\begin{align}
\dot{P}_{x_0\bar{x}}(t) =  \frac{d}{dt} \int_{0}^t \big( \frac{a^2}{2} \frac{M(t-t') P_{x_0\bar{x}+1}(t') }{a^2} -\\
\frac{-2M(t-t')P_{x_0\bar{x}}(t') + M(t-t') P_{x_0\bar{x}-1}(t') dt'}{a^2} \big)
\end{align}
and then to the form of differential equation % with continuous function $M(t)$
\begin{align}
\frac{d }{d t}P_{x_0x}(t) = \frac{a^2}{2}\frac{d^2}{d x^2} \frac{d}{dt} \int_{0}^{t}M(t-t')P_{x_0x}(t')dt' , 
\label{eq_fin_gme}
\end{align}% D(x) D P(x,t) 
where $a$ denotes a distance between two neighboring sites of one dimensional lattice. % which we later can put to $1$ without loss of generality.  

(i) The case when all \emph{travel times are exponential} with parameter $\tau_{\bar{x}}$ 
   $\tilde{\psi}_{\bar{x}}(s) =1/(1+s\tau_{\bar{x}})$ gives us then 
   \begin{align}
\tilde{M}_{\bar{x}}(s) =\frac{1}{s\tau_{\bar{x}}}.
   	\label{eq_m_exp_first}
   \end{align}
Assuming continuous dependence of $\tau_{\bar{x}}$ parameter from $\bar{x}$ and putting $D(\bar{x}) = \frac{a^2}{2\tau_{\bar{x}}}$, 
 Eq.~(\ref{eq_dif_lapl}) can be transformed to a continuous form of a diffusion equation in the Ito form, which coincides with the results obtained in \cite{Barkai,Landman1977}
\begin{align}
\frac{d }{d t}P_{x_0x}(t) = \frac{d^2}{d x^2} D(x)P_{x_0x}(t) , %\green{\frac{a^2}{2}}
\label{eq_ito_form}
\end{align}
where $D(x) = \lim_{\bar{x}\rightarrow x}\frac{a^2}{2\tau_{\bar{x}}}$. %for continuous $\tau_x$  % make simulations for $\tau$ depending on x coordinate  
 We note that this configuration mimics the kinetics of the trap model with exponential waiting times \cite{Sokolov2010}.
The first interpretation is closely related to continuous limits for Markov jump processes \cite{Stroock}.

(ii) In the case when  \emph{all mean travel times are finite} but not necessarily exponential, one can write the  second order expansion valid for small $s$:
$\tilde{\psi}_{\bar{x}}(s) = 1- s\langle\tau_{\bar{x}}\rangle - s^2/2\langle\tau_{\bar{x}}\rangle^2+  o(s^2)$.
Then the function $\tilde{M}_{\bar{x}}(s)$ is simply
\begin{align}
\tilde{M}_{\bar{x}}(s) \propto \frac{1-s\langle\tau_{\bar{x}}\rangle -s^2/2\langle\tau_{\bar{x}}\rangle^2}{s\langle\tau_{\bar{x}}\rangle +s^2/2\langle\tau_{\bar{x}}\rangle^2},
\label{eq_mx_s_exp}
\end{align}
which gives the correction in the real time domain $-\frac{1}{\langle \tau_{\bar{x}}\rangle} + e^{-t/\langle\tau_{\bar{x}}\rangle} +\delta(t)$, and differs from the terms in Eq.~(\ref{eq_m_exp_first}). %where both denominator and divisor contain more terms.
Then from  
Eq.~(\ref{eq_dif_lapl})  we get:  %substituting $M_{\bar{x}}(s)$ from Eq.~(\ref{eq_mx_s_exp}) gives 
\begin{align}
s\tilde{P} _{x_0\bar{x}}(s) -P_{x_0\bar{x}}(0) = \\ \nonumber \sum_{x'}Q_{\bar{x}x'}\Big(  s\tilde{M}_{x'}(s) \tilde{P}_{x_0x'}(s) \Big) - 
s\tilde{M}_{\bar{x}}(s) \tilde{P}_{x_0\bar{x}}(s).
\end{align}
We can transform this to the form of differential equation with the integro-differential operators
\begin{align}
\frac{d }{d t}P_{x_0x}(t) = \frac{a^2}{2}\frac{d^2}{d x^2} \frac{d}{dt} \int_{0}^{t}M_x(t-t')P_{x_0x}(t')dt'.
\label{eq_gen_ito}
\end{align}
The main difference between differential equations (\ref{eq_fin_gme}) and (\ref{eq_gen_ito}) is that in case (o) when $\psi_{\hat{x}}(t)=\psi(t)$ the function $M_{x}(t)$ is site-independent. %\blue{(we can also express $D(\bar{x})$  according to the  $\tilde{\psi}_{\bar{x}}(s)$ small $s$-expansion)}
Eq.~(\ref{eq_gen_ito}) has more general form than the diffusion equation for homogeneous CTRW \cite{Barkai}.

(iii) The last case is 
when \emph{at least one mean travel time is infinite.}
First we put all travel time distributions to be power laws with  exponents coordinate-dependent $\alpha(x)$: $\psi_{\bar{x}}(t) = t^{-1-\alpha(\bar{x})}$, we get 
\begin{align}
M_{\bar{x}}(t) \propto \frac{1}{\langle\tau_{\bar{x}}\rangle}t^{-1+\alpha(x)}-\delta(t).
\end{align}
Then the diffusion equation has the integro-differential operators with parameter $\alpha$:
\begin{align}
\frac{d P_x(t)}{dt} = \frac{d^2}{dx^2} (D(x)  \textfrak{D}_t^{1-\alpha(x)}P_x(t)),
\end{align}
where  $ \textfrak{D}^{1-\alpha(x)}_t$ is the generalization of the Riemann-Liouville derivative of order $1-\alpha(x)$: %($\alpha(x)$ depends on $x$), defined as follows
\begin{align}
\textfrak{D}^{1-\alpha(x)}_t (P_{x_0x}(t) ) = \frac{1}{\Gamma(1-\alpha(x))}\frac{d}{dt} \int_0^t  \frac{P_{x_0x}(t')dt' }{(t-t')^{\alpha(x)}}.
\end{align}
%(change notation for $D(x)$ diffusion coefficient and define operator)  We also need to consider 
When all $\tilde{\psi}_{\bar{x}}(s)$ exhibit power law behaviour with the same scaling exponent $\alpha$:  
$\tilde{\psi}_{\bar{x}}(s) = 1- s^{\alpha}\langle\tau_{\bar{x}}\rangle^{\alpha} +o(s^{\alpha})$, 
the function $\tilde{M}_{\bar{x}}(s)$ is: %\blue{make second order correction}
\begin{align}
\tilde{M}_{\bar{x}}(s) \propto \frac{1-s^{\alpha}\langle\tau^{\alpha}_{\bar{x}}\rangle}{s^{\alpha}\langle\tau_{\bar{x}}\rangle^{\alpha}},
\label{eq_mx_s_expalp}
\end{align}
which gives the differential form of the equation as in \cite{Chechkin2006}.
%\begin{align}
%M_{\bar{x}}(s) \propto \frac{1-s^{\alpha}\langle\tau^{\alpha}_{\bar{x}}\rangle -s^{2\alpha}/2\langle\tau_{\bar{x}}\rangle^{2\alpha}}{s^{\alpha}\langle\tau_{\bar{x}}\rangle^{\alpha} +s^{2\alpha}/2\langle\tau_{\bar{x}}\rangle^{2\alpha}}.
%\label{eq_mx_s_expalp}
%\end{align}

%\subsubsection{Specific example for the first interpretation} \label{subsec_first}
We consider the specific example of the equation with exponential travel times $\tilde{\psi}_x(s)=1/(1+s\tau(x))$ and temporal heterogeneities introduced $\tau(x)=\sin(x)$, as it was considered for the interval in \cite{GrebTupikina}. 
Then  Eq.~(\ref{eq_ito_form}) takes a form 
 \begin{align}
 \frac{d }{d t}P_{x_0x}(t) = \frac{d^2}{d x^2}\big( \frac{a^2}{2\sin(x)}P_{x_0x}(t) \big)
 \label{eq_sdf}
 \end{align}
 with the space-dependent diffusion coefficient. %(compare the difference with the third interpretation, where diffusion coefficient become space- and time-dependent, see  \ref{subsec_third}). %For equation above we can also get analytic solution. \red{Solution for \ref{eq_sdf} can be found using the subordinated processes \cite{Weron}.}

\subsection{Second HCTRW interpretation}
\label{sec_deriv2}

In the second and third HCTRW interpretations the waiting time distribution $\tilde{\psi}_{\bar{x}}(s)$ depends on  neighboring nodes of $\bar{x}$, while in the first interpretation $\tilde{\psi}_{\bar{x}}(s)$ depends on $\bar{x}$ only, section \ref{subsec_1interpr}.  
For the second and third interpretations the general form of GME
Eq.~(\ref{eq_dif_lapl}) can be simplified:
\begin{align}
s\tilde{P}_{x_0\bar{x}}(s)-P_{x_0\bar{x}}(0) = \\ \nonumber 
s(\tilde{M}_{\bar{x}+1}(s) \tilde{P}_{x_0\bar{x}+1}(s) - \tilde{M}_{\bar{x}}(s) \tilde{P}_{x_0\bar{x}}(s)) + \\ \nonumber
s(\tilde{M}_{\bar{x}-1}(s) \tilde{P}_{x_0\bar{x}}(s) - \tilde{M}_{\bar{x}}(s) \tilde{P}_{x_0\bar{x}}(s)).
\label{eq_exp_mxs}
\end{align}
%from which we get: \begin{align}\tilde{P}_{x_0\bar{x}}(s)-P_{x_0\bar{x}}(0) = \\ \nonumber s\big(M_{\bar{x}}(s) (\tilde{P}_{x_0\bar{x}+1}(s)-\tilde{P}_{x_0\bar{x}}(s))\big)  \\ \nonumber +s\big(M_{\bar{x}}(s) (\tilde{P}_{x_0\bar{x}-1}(s)-\tilde{P}_{x_0\bar{x}}(s))\big) +\\ \nonumber s\big((M_{\bar{x}+1}(s) -M_{\bar{x}}(s))\tilde{P}_{x_0\bar{x}+1}(s)   \\ \nonumber + (M_{\bar{x}-1}(s) -M_{\bar{x}}(s))\tilde{P}_{x_0\bar{x}-1}(s)\big). \end{align}%where we grouped components in order to separate two parts on the right-hand side of the equation:
%\begin{align} s\tilde{P}_{x_0\bar{x}}(s)-P_{x_0\bar{x}}(0) = \\ \nonumber  a^2s\frac{M_{\bar{x}}(s) (\tilde{P}_{x_0\bar{x}-1}(s)-2\tilde{P}_{x_0\bar{x}}(s)) +\tilde{P}_{x_0\bar{x}+1}(s)}{a^2} +  \\ \nonumber a^2s \frac{(M_{\bar{x}+1}(s) -M_{\bar{x}}(s))\tilde{P}_{x_0\bar{x}+1}(s)}{a^2}\\ \nonumber +a^2s\frac{(M_{\bar{x}-1}(s)-M_{\bar{x}}(s))\tilde{P}_{x_0\bar{x}-1}(s)}{a^2}, \end{align}
Then after simplifying 
the right-hand side of the equation and taking the limit we come to the expression:
\begin{align}
\lim_{a\rightarrow 0} 
\big(
\frac{(\tilde{M}_{\bar{x}+1}(s) -\tilde{M}_{\bar{x}}(s))\tilde{P}_{x_0\bar{x}+1}(s)}{a^2}
\\ \nonumber	
- \frac{(\tilde{M}_{\bar{x}}(s) -\tilde{M}_{\bar{x}-1}(s))\tilde{P}_{x_0\bar{x}-1}(s)}{a^2} \big) = \\ \nonumber 
\lim_{a\rightarrow 0}  \frac{\tilde{M}_{\bar{x}+1}(s) -\tilde{M}_{\bar{x}}(s)}{a} 
\lim_{a\rightarrow 0} \frac{\tilde{P}_{x_0\bar{x}+1}(s) -\tilde{P}_{x_0\bar{x}-1}(s)}{a}.
\end{align}
Assuming a slow change of $\tilde{M}_{\bar{x}}(s)$ from $\bar{x}$ we can put:  $\tilde{M}_{\bar{x}+1}(s) -\tilde{M}_{\bar{x}}(s) \approx \tilde{M}_{\bar{x}}(s) -\tilde{M}_{\bar{x}-1}(s)$. 
 Then the evolution equation is:
\begin{align}
s\tilde{P}_{x_0{x}}(s)-P_{x_0{x}}(0) = \\ \nonumber
a^2s \tilde{M}_x(s) \frac{d^2 P_{x_0x}(s) }{dx^2} + a^2s\Big(\frac{d \tilde{M}_x(s)}{dx}\Big) \Big(\frac{d P_{x_0x}(s)}{dx}\Big).
%\label{eq_p_s_m_s}
\end{align}
Then the right-hand side can be further transformed to 
\begin{align}
s\tilde{P}_{x_0{x}}(s)-P_{x_0{x}}(0) =\frac{d}{dx}\tilde{M}_x(s)\frac{d}{dx}\tilde{P}_{x_0x}(s) .
\label{eq_dif_m_xs}
\end{align}
 %\blue{We aim to get expression for  $D(x)$ in terms of $M_x(s)$ functions: (and find what is completely different from I interpretation)} i.e. %\blue{(to correct)} % \blue{ (H\"anggi formalism)}.$D(x) =sM_x(s) $ in particular case. 
Note that Eq.~(\ref{eq_dif_m_xs})  differs from  Eq.~(\ref{eq_ito_form}), obtained using  general assumptions about $M_{\bar{x}}(t)$ properties and allowing a waiting time distribution to depend on $\bar{x},x'$ nodes. 

\subsubsection{Second  continuous HCTRW interpretation in one dimension}
\label{sec_second}

Here we consider \emph{the second continuous HCTRW interpretation} for one-dimensional case. % (for simplicity),
In this case a random walk spends in each site $\bar{x}$ time driven from the distribution $\psi_{\bar{x}x'}(t)$, which 
depends on both $\bar{x}, x'$, hence in this case we can not use the assumption of locality as in the first interpretation. 
 When a travel time distribution can be represented  in a form of linear combination of functions $\psi_{\bar{x}x'}(t) = \gamma_1\psi_{\bar{x}}(t) +\gamma_2\psi_{x'}(t)$ for some parameter $\gamma$, this case can be analyzed using the first and third interpretations. 
%This interpretation can have the most general form,  when $\psi_{\bar{x}x'}(t) $ is not  represented as  	the combination of $\psi_{\bar{x}}(t), \psi_{x'}(t)  $ (see calculations for case $\psi_{\bar{x}x'}(t) =a\psi_{\bar{x}}(t)+ b\psi_{x'}(t) $, based on in Appendix in \cite{GrebTupikina}). 
For the symmetric case with equal $Q_{\bar{x}x'}$ for all $x'$ and fixed $\bar{x}$ we simply can write 
%\begin{align}M_{\bar{x}}(s) = \frac{\sum_{{x}'} Q_{\bar{x}x'}\tilde{\psi}_{\bar{x}x'}(s)}{1-\sum_{{x}'}Q_{\bar{x}x'}\tilde{\psi}_{\bar{x}x'}(s)}, \label{eq_m_x_gen} \end{align}
\begin{align}
\tilde{M}_{\bar{x}}(s) = \frac{\sum_{{x}' =\bar{x}\pm 1}\tilde{\psi}_{\bar{x}x'}(s)}{2-\sum_{{x}'\bar{x}\pm 1}\tilde{\psi}_{\bar{x}x'}(s)}.
\end{align}

Further we use the assumption of smoothness of $M_{\bar{x}}(t)$. %see Appendix \ref{sec_append2}. 
 We refer to calculations for the third interpretation in Subsection \ref{sec_third} and define new functions $M^{\bar{x}+1}_{\bar{x}}(t)$,  $M_{\bar{x}}(t)$. 
Coming to the continuous limit in Laplace domain 
\begin{align}
s\tilde{P}_{x_0x}(s)- {P}_{x_0x}(0)=
\frac{d}{d x} s M_x(s)\frac{d}{d x} \tilde{P}_{x_0x}(s),
% \partial_x \frac{d}{dt} \int_{0}^t M_x(t-t') \partial_x P_{x_0x}(t') dt'.
\label{eq_dif_general_s}
\end{align}
the right-hand side of which gives 
$D(x) \tilde{P}''_{x_0x}(s) + D'(x) \tilde{P}'_{x_0x}(s)$.
Comparing Eq.~\ref{eq_dif_m_xs} with Eq.~(\ref{eq_dif_general_s}) we deduce some specific properties of the second interpretation.
Now as in the first interpretation we will consider several different cases for the travel time types.

(o) We start with the case 
 when all mean travel times of $\tilde{\psi}_{\bar{x}x'}(t)$  are finite $\langle\tau_{\bar{x}x'} \rangle$:
 $ \tilde{\psi}_{\bar{x}x'}(s) = 1-s \langle\tau_{\bar{x}x'} \rangle + \frac{s^2}{2}\langle\tau_{\bar{x}x'} \rangle^2 +o(s^2)$ for small $s$. %including the second order term as in Eq.~(\ref{eq_mx_s_exp}). 
When taking into account only first order terms we get: % the second interpretation for continuous  HCTRW:
\begin{align}  
M_{\bar{x}}(s) % \approx    \frac{\frac{1}{2}(2-s(\langle\tau_{\bar{x}\bar{x}+1} \rangle +\langle\tau_{\bar{x}\bar{x}-1} \rangle) }{1-\frac{1}{2}(2-s(\langle\tau_{\bar{x}\bar{x}+1} \rangle +\langle\tau_{\bar{x}\bar{x}-1} \rangle))}
%= \\ \nonumber
 =\frac{1-s\hat{\tau}_{\bar{x}}}{s\hat{\tau}_{\bar{x}}} ,
\end{align}
where we call $\hat{\tau}_{\bar{x}} =(\langle\tau_{\bar{x}\bar{x}+1} \rangle +\langle\tau_{\bar{x}\bar{x}-1} \rangle)/2$.  %(symmetry of  $\langle\tau_{\bar{x}\bar{x}-1} \rangle$ in respect to $\bar{x}$ indices is not yet considered here). 
In this case then we get $M_{\bar{x}}(t) \approx 1/\hat{\tau}_{\bar{x}} - A(\hat{\tau}_{\bar{x}}) \delta(t) $,  
where coefficient  $A(\hat{\tau}_{\bar{x}}) $ is set by higher order terms.  
Since $\hat{\tau}_{\bar{x}}$  depends on the neighbouring sites, we can use the assumption of slowly changing $M_{\bar{x}}(s)$ from $\bar{x}$. %(which in this case does not depend on $t$). since then we get memoryless process
Therefore  substituting these functions to Eq.~(\ref{eq_dif_lapl})  and using Taylor series we come to the continuous limit:
%\begin{align} s\tilde{P}_{x_0\bar{x}}(s) -P_{x_0\bar{x}}(0) = \\ \nonumber \sum_{x'=\bar{x}\pm 1}\frac{1}{2}\Big(  s \frac{1-s\hat{\tau}_{x'}}{s\hat{\tau}_{x'}}  \tilde{P}_{x_0x'}(s) \Big) - s  \frac{1-s\hat{\tau}_{\bar{x}}}{s\hat{\tau}_{\bar{x}}} \tilde{P}_{x_0\bar{x}}(s). \end{align}
\begin{align}
s\tilde{P}_{x_0\bar{x}}(s) -P_{x_0\bar{x}}(0) = \\ \nonumber
\sum_{x' =
	\bar{x}\pm 1}\frac{1}{2}\Big(   \frac{1-s\hat{\tau}_{x'}}{\hat{\tau}_{x'}}  \tilde{P}_{x_0x'}(s) \Big) - 
 \frac{1-s\hat{\tau}_{\bar{x}}}{\hat{\tau}_{\bar{x}}} \tilde{P}_{x_0\bar{x}}(s).
 \label{eq_cont_int_two}
\end{align}

This is the "stepping stone"  of the derivation for the second continuous HCTRW interpretation. 
%The diffusion coefficient is calculated as the continuous limit of the expression $D(\bar{x}) = F(\hat{\tau}_{\bar{x}})$.
We compare the diffusion coefficient  from Eq.~\ref{eq_cont_int_two} with the first interpretation,  Eq.~(\ref{eq_m_exp_first}) with $D(\bar{x}) = \frac{a^2}{2{\tau}_{\bar{x}}}$.

% \textbf{ Let us now consider the Stratonovich formalism}.  
%Applying the same analogy as for the Ito formalism, we can find the relation of function $g(x,t)$ with  $M_x(s)$, which for 1D is expressed through: 
%\begin{align}
%\frac{d g(x,t)}{d x} \frac{d g(x,t) P_{x_0x}(s)}{dx} =\green{ \frac{d^2}{dx^2} sM_x(s) P_{x_0x}(s)}
%\end{align}

\subsubsection{Second interpretation with exponential travel time distributions}
Here we consider particular case of the second interpretation with the exponential travel time distribution.
Similar case was considered in \cite{Bouchard}, where transition rates $Q_{\bar{x}\bar{x}+1}$ have two parameters $D_{\bar{x}\bar{x}+1}$ and $F_{\bar{x}\bar{x}+1}$  %(not with only one as we usually had for $\psi_{\bar{x}\bar{x}+1}(t)$).
 We note that here we consider ME and calculations for GME will be done further.
When the rate parameters depend on the starting and end points, then: 
\begin{align}
Q_{\bar{x}\bar{x}+1} = \frac{D_{\bar{x}\bar{x}+1} }{a^2}\exp(-\frac{aF_{\bar{x}\bar{x}+1}}{2\gamma D_{\bar{x}\bar{x}+1}}), 
\end{align}
where $F_{\bar{x}\bar{x}+1}$ is microscopic parameter, related to the interaction with the medium, $a$ is the parameter of the lattice. 
Using the expansion in powers of $a$ we get:
\begin{align}
\frac{d}{dt} P_{x_0\bar{x}+1} \approx \\ \nonumber \frac{1}{a}\big(D_{\bar{x}\bar{x}+1} \frac{P_{x_0\bar{x}+1}-P_{x_0\bar{x}}}{a} -D_{\bar{x}-1\bar{x}} \frac{P_{x_0\bar{x}}-P_{x_0\bar{x}-1}}{a}\big) \\ \nonumber
+\frac{1}{a}\big(\frac{F_{\bar{x}-1\bar{x}}}{\gamma} \frac{P_{x_0\bar{x}+1}-P_{x_0\bar{x}}}{2} -\frac{F_{\bar{x}\bar{x}+1}}{\gamma} \frac{P_{x_0\bar{x}}-P_{x_0\bar{x}-1}}{2} \big)  \\ \nonumber 
+ \frac{1}{8}\big( \frac{F^2_{\bar{x}\bar{x}+1}}{\gamma^2D_{\bar{x}\bar{x}+1}} (P_{x_0\bar{x}+1}-P_{x_0\bar{x}}) \\ \nonumber -\frac{F^2_{\bar{x}-1\bar{x}}}{\gamma^2D_{\bar{x}\bar{x}-1}} (P_{x_0\bar{x}}-P_{x_0\bar{x}-1}) \big)
\label{eq_gme_bou}
\end{align}
Then in the limit we get Fokker-Planck in the form 
\begin{align}
\frac{d}{dt} P_{x_0x}(t) = %\blue{\frac{d}{dx} D(x)\frac{d}{dx}P} + 
\frac{d}{dx} (-\frac{F(x)}{\gamma}P +D(x) \frac{d}{dx} P),
\end{align}
where we  $D(x) = \lim_{\bar{x}\rightarrow x} \frac{a^2}{\tau_{\bar{x}+1} +\tau_{\bar{x}-1}}$.
%which hence corresponds to $\alpha=0$, or Ito convention.

In the case when the rate microscopic parameter $D$ depends only on one of the microscopic evaluation points, as in the asymmetric case \cite{Bouchard}, e.g.: 
\begin{align}
Q_{\bar{x}\bar{x}+1} = \frac{D_{\bar{x}+1} }{a^2}\exp(-\frac{aF_{\bar{x}\bar{x}+1}}{2\gamma D_{\bar{x}+1}}), \\ 
Q_{\bar{x}+1\bar{x}} = \frac{D_{\bar{x}} }{a^2}\exp(-\frac{aF_{\bar{x}\bar{x}+1}}{2\gamma D_{\bar{x}}}).
\end{align}
From this we get
\begin{align}
\frac{d}{dt} P_{x_0\bar{x}+1} \approx \frac{1}{a}\big(D_{\bar{x}+1} \frac{P_{x_0\bar{x}+1}-P_{x_0\bar{x}}}{a} -D_{\bar{x}} \frac{P_{x_0\bar{x}}-P_{x_0\bar{x}-1}}{a}\big) +\\ \nonumber
+\frac{1}{a}\big(\frac{F_{\bar{x}-1\bar{x}}}{\gamma} \frac{P_{x_0\bar{x}+1}-P_{x_0\bar{x}}}{2} -\frac{F_{\bar{x}\bar{x}+1}}{\gamma} \frac{P_{x_0\bar{x}}-P_{x_0\bar{x}-1}}{2} \big) + \\ \nonumber 
+ \frac{1}{8}\big( \frac{F^2_{\bar{x}\bar{x}+1}}{\gamma^2D_{\bar{x}+1}} (P_{x_0\bar{x}+1}-P_{x_0\bar{x}}) -\frac{F^2_{\bar{x}-1\bar{x}}}{\gamma^2D_{\bar{x}}} (P_{x_0\bar{x}}-P_{x_0\bar{x}-1}) \big).
\label{eq_gme_2}
\end{align} 
Then the corresponding Fokker-Planck should give the form 
\begin{align}
\frac{d}{dt} P(x,t) = \frac{d}{dx} (-\frac{F(x)}{\gamma}P(x,t) +\frac{d}{dx}D(x) P(x,t)),
\end{align}
which hence corresponds to another diffusion convention, the Stratonovich form for diffusion, related to kinetic interpretation, as noted in  \cite{Sokolov2010}.
%{We get the Stratonovich form for the diffusion, the reason for this is coupled to the way of the equation derivation.}
Moreover, for different discrete models, the diffusion term can be written in the form:
\begin{align}
\frac{d}{dx} (D(x)^\alpha \frac{d}{dx}D(x)^{1-\alpha} P),
\end{align}
where parameter $\alpha\in [0,1].$
%Now the question is how to use the same mapping for denoting $\psi_{\bar{x}x'}(t) $ as $ \exp(\frac{D_{\bar{x}x'}}{F})$ or as , or as $\exp $.

\subsection{Third HCTRW interpretation}
\label{sec_third}

Finally, we  consider \emph{the third  continuous interpretation of HCTRW}, when  
each travel time distribution $\psi_{\bar{x}x'}(t)$ depends only on the end point $x'$, where the random walk jumps. 
For convenience, we first consider one-dimensional case and put the transition matrix $Q_{\bar{x}x'} = 1/2$.
Then the expression for the function $M_{\bar{x}}(s)$ is expressed explicitly as the function of the neighboring sites of $\bar{x}$, using the derivations from Subsection \ref{sec_deriv}:
\begin{align}
M_{\bar{x}}(s) = \frac{\sum_{x'=\bar{x}\pm 1}\tilde{\psi}_{x'}(s)}{2-\sum_{x'=\bar{x}\pm 1}\tilde{\psi}_{x'}(s)}.
\end{align}

When all \emph{travel time distributions are exponentials}, we get: 
\begin{align}
M_{\bar{x}}(s) = \frac{s(\tau_{\bar{x}+1}+\tau_{\bar{x}-1})+2}{s(\tau_{\bar{x}+1}+\tau_{\bar{x}-1})+2s^2\tau_{\bar{x}+1}\tau_{\bar{x}-1}},
\label{eq_mx_exp}
\end{align}
where kernel $M_{\bar{x}}(s)$  depends on local properties of neighboring nodes of $\bar{x}$. 
Inserting the Taylor series expansion for Eq.~(\ref{eq_mx_exp}) we regroup components to see the difference with Eq.~(\ref{eq_m_exp_first})
\begin{align}
M_{\bar{x}}(s) = \frac{1}{s\tau_{\bar{x}}} (1 + s\delta \phi_{\bar{x}}) (1-s\delta \phi_{\bar{x}}) (1-\frac{a^2 \phi ''}{2 \phi_{\bar{x}}}),
\end{align}
where we denoted $\tau_x = \delta \phi_x$ so that later we can use the standard notations $D = \frac{a^2}{2\delta}$.
\begin{comment}
\begin{align}
M_{\bar{x}}(s) = \frac{1}{s\tau_{\bar{x}}} + \frac{\mathcal{B}(a,s)}{\mathcal{A}(a,s)},
\end{align}
 where 
 \begin{align}
 \mathcal{A}(a,s) = 1+ s(\frac{a^2\tau''}{2} +s\tau_{\bar{x}}T(a) + \frac{a^2\tau''}{2}T(a)), 
 \\
 \blue{
 \mathcal{B}(a,s) = \frac{s(\tau_{\bar{x}} +\frac{a^2\tau''}{2})-s(\frac{a^2\tau''}{2} +s\tau_{\bar{x}}T(a) + \frac{a^2\tau''}{2}T(a))}{s\tau_{\bar{x}}} },
 \\
 T(a) = \frac{\tau_{\bar{x}}^2 -a^2\tau'^2}{\tau_{\bar{x}} +a^2\tau''/2}.
 \end{align}
  This allows us to come to the limit for fixed diffusion constant $D=a^2/\delta$, where $\delta = \tau_{\bar{x}}/t(\bar{x})$.
  \begin{align}
M_{\bar{x}}(s) = \frac{1}{s\tau_{\bar{x}}} +
 \frac{1-T(a)(1 +\frac{a^2\tau''}{2\tau_{\bar{x}}})}{1+s(\frac{a^2\tau''}{2} +s\tau_{\bar{x}}T(a) + \frac{a^2\tau''}{2}T(a))}.
\end{align}
\end{comment}
Directly from Eq.~(\ref{eq_mx_exp}) %Eq.~(\ref{eq_exp_mxs}) 
we get 
\begin{align}
M_{\bar{x}}(t) = \frac{2}{\tau_{\bar{x}+1}+\tau_{\bar{x}-1}} + \\ \nonumber e^{-t\frac{\tau_{\bar{x}+1}+\tau_{\bar{x}-1}}{\tau_{\bar{x}+1}\tau_{\bar{x}-1}}} \frac{\tau_{\bar{x}+1}-\tau_{\bar{x}-1}}{2\tau_{\bar{x}+1}\tau_{\bar{x}-1}(\tau_{\bar{x}+1}+\tau_{\bar{x}-1})},
\end{align} 
where in the limit the second term on the right hand side vanishes to zero. 
 The expansion of $\tau_x$ gives us:
 $
 \tau_{\bar{x}+1} \approx \tau_{\bar{x}} + a \tau'_{\bar{x}} + a^2/2 \tau''_{\bar{x}}$. % assuming differentiability.
%hence to simplify expressions with convolution we derive further equations in Laplace domain.
Using the expression for the function $M_x(t)$ from Eq.~(\ref{eq_mx_exp}) we get  the equation with two convolutions in the real time domain:
  \begin{align}
\frac{d P_{x_0x}(t)}{dt} = M_x(t) P_{x_0x}''(t) + M_x'(t) P_{x_0x}'(t).
  \end{align}
Note that for the first interpretation of continuous HCTRW the function $M_x = \lim_{\bar{x}\rightarrow x} \frac{a^2}{2\tau_{\bar{x}}}$ for the exponential travel time distributions, Eq.~(\ref{eq_ito_form}). %\blue{(commutativity of $(\bar{x},t)\rightarrow(\bar{x},s)\rightarrow (x,s)\rightarrow (x,t)$)}.
We regroup components so that the final expression becomes:
\begin{align}
\frac{dP_{x_0x}(t)}{dt} = \frac{dD(x)}{dx}  \frac{dP_{x_0x}(t)}{dx} ,
\end{align}
where the right hand side transforms to $ D(x) P_{x_0x}''(t) + D'(x) P_{x_0x}'(t)$.
For the exponential travel times $ D(x) = a^2/(2\tau(x))$.
Or in Laplace domain: 
\begin{align}
s\tilde{P}_{x_0x}(s) - P_{x_0x}(0) = \frac{dD(x)}{dx}  \frac{d\tilde{P}_{x_0x}(s)}{dx} ,  
\end{align}
the right-hand side of which can be expressed  as $ D(x) \tilde{P}''_{x_0x}(s) + D'(x) \tilde{P}'_{x_0x}(s)$.
Different way to calculate the continuous limits of GME  is using the Taylor series of functions from Eq.~(\ref{eq_dif_lapl}):
$$\psi_{\bar{x}\pm1}(t) = \psi_{\bar{x}}(t) \pm a\psi_{\bar{x}}(t)' +a^2/2\psi_{\bar{x}}(t)'' ,$$
where $a$ is a lattice parameter. Then
\begin{align}
\tilde{P}_{x_0 \bar{x}+1}(s) \approx \tilde{P}_{x_0\bar{x}}(s) + a \tilde{P}'_{x_0\bar{x}}(s) + \frac{a^2}{2} \tilde{P}''_{x_0\bar{x}}(s), %o(a^21),
\label{eq_p_tayl}
\end{align}
where higher order terms can be neglected.
Then we regroup components in Eq.~(\ref{eq_dif_lapl}) such that the first and second derivative in $x$ of $\tilde{P}_{x_0x}(s)$ from Eq.~(\ref{eq_p_tayl}) are separated.
We substitute Taylor expansion for functions in Eq.~(\ref{eq_dif_lapl}) to get for the one-dimensional symmetric case ($Q_{\bar{x}\pm 1}=1/2$):
\begin{align}
s\tilde{P}_{x_0\bar{x}}(s) -\tilde{P}_{x_0\bar{x}}(0) = 
\\ \nonumber
\frac{1}{2}\sum_{x' =\bar{x}\pm 1}\Big(  sM_{x'}(s) \tilde{P}_{x_0x'}(s) \Big) - 
sM_{\bar{x}}(s) \tilde{P}_{x_0\bar{x}}(s).
\label{eq_p_tayl_s}
\end{align}
The right-hand side after regrouping components gives
\begin{align}
\frac{1}{2}\Big(  sM_{\bar{x}+1}(s) (P_{x_0\bar{x}}(s) +aP_{x_0\bar{x}}'(s) +a^2/2 P_{x_0\bar{x}}''(s)) 
+ \\ \nonumber
 sM_{\bar{x}-1}(s)(P_{x_0\bar{x}}(s) -aP_{x_0\bar{x}}'(s) +a^2/2 P_{x_0\bar{x}}''(s)) 
 \Big) - \\ \nonumber
-sM_{\bar{x}}(s) P_{x_0\bar{x}}(s)
=
\\ \nonumber
\frac{1}{2}\Big(  sM_{\bar{x}+1}(s) (P_{x_0\bar{x}}(s) +aP_{x_0\bar{x}}'(s) +a^2/2 P_{x_0\bar{x}}''(s)) 
-\\ \nonumber
-sM_{\bar{x}}(s) P_{x_0\bar{x}}(s)) 
+ \\ \nonumber
(sM_{\bar{x}-1}(s)(P_{x_0\bar{x}}(s) -aP_{x_0\bar{x}}'(s) +
\\ \nonumber a^2/2 P_{x_0\bar{x}}''(s)) 
- 
sM_{\bar{x}}(s) P_{x_0\bar{x}}(s)\Big) .
\end{align}
%Now we substitute Taylor series for $M_{\bar{x}\pm 1}$.
Then we transform Eq.~(\ref{eq_p_tayl_s}) by combining together the derivatives of the same order:
\begin{align}
s\tilde{P}_{x_0\bar{x}}(s) -\tilde{P}_{x_0\bar{x}}(0) = \\ \nonumber
A_{\bar{x}} (s) P_{x_0\bar{x}(s)} + B_{\bar{x}}(s) P_{x_0\bar{x}}'(s) + C_{\bar{x}}(s)  P_{x_0\bar{x}}''(s),
\end{align}
where each coefficient is expressed as follows:
\begin{align}
A_{\bar{x}}(s) = \frac{a^2s}{2}M_{\bar{x}}''(s), \\
B_{\bar{x}}(s) = a^2s M_{\bar{x}}'(s), \\
 C_{\bar{x}}(s) = \frac{a^2s}{2}(M_{\bar{x}}(s) + \frac{a^2}{2}M_{\bar{x}}''(s)),
\end{align}
where from the last equation for $C$ we eliminate later the component $\frac{a^2}{2}M_{\bar{x}}''(s)$. %(\blue{see notes}).

We also note that the  HCTRW  with travel times $\psi_{\bar{x}x'}(t) = \psi_{x'}(t)$ can be considered as the limiting case of the second interpretation for $\alpha\rightarrow 0$ in  $\psi_{\bar{x}x'}(t) = \psi(\alpha \bar{x} +(1-\alpha)x',t)$. 
Then by making travel times to be driven from exponential distributions, 
the HCTRW can be mapped to the so-called accordion model \cite{Sokolov2010}.

\begin{figure}
	\includegraphics[width=2.7in]{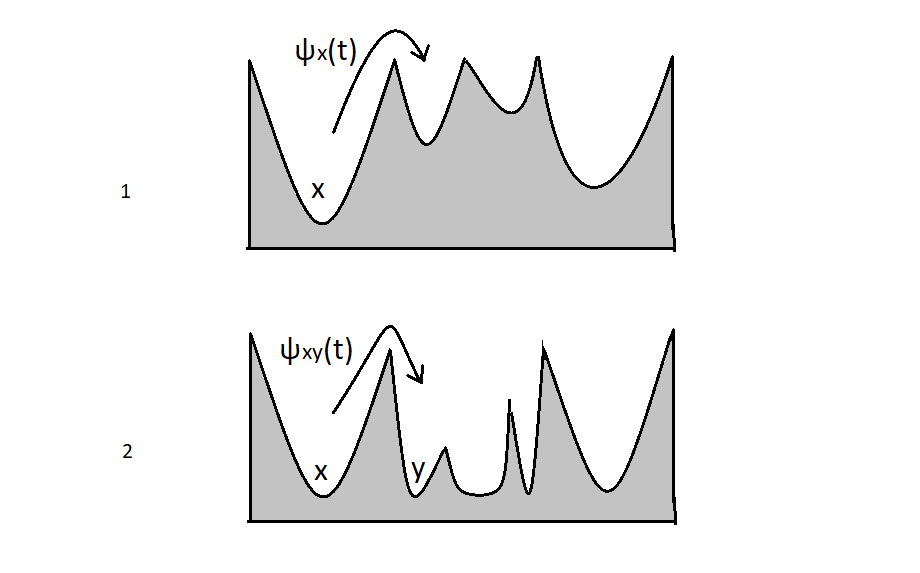}\centering
	\caption{Illustration of connection between the HCTRW model an the  trap model (1) or barrier model (2). }	
	\label{fig_barri}
\end{figure}

%In HCTRW model the random walk jumps with travel times $\psi_{xx'}$ from site $x$ to site $x'$\green{alike Stratonovich}. From such HCTRW model we can make another process, the model where random walk jumps with travel times $\psi_x$ depending only on the initial node \green{(alike Ito, evaluation only in the beginning of the process).}

\section{Discussions}
\label{sec_disc}
%\subsection{Ito versus Stratonovich conventions}
In this paper we derived studied the continuous limits of heterogeneous continuous time random walk model (notation used in the text is HCTRW). We  derived the generalized master equation for the HCTRW model and considered. 
%We also compare our results with the GME for other stochastic models. 

%\blue{(What has been done before?)} 
Previously, continuous limits of random walk models were considered for various models of stochastic processes.
Dynamics of homogeneous CTRW model on a lattice and continuous limits for homogeneous CTRW on lattices were considered in \cite{Chechkin2006}. 
 In \cite{Angstmann2015} the Generalized Master Equation (GME) for discrete time random walk was considered for various random walk models.  
In \cite{Lambiotte2011} the framework for continuous limits for CTRW model was developed. %In \blue{(arxiv rev from Appl.Netw.Journ.)}  
The diffusion coefficient, in general, depends on a combination of two kinetic parameters \cite{Sokolov2010}: mean free path and correlation time. Using parametrisation as in \cite{Sokolov2010} one can encode diffusion in inhomogeneous medium. 
 %(make comparison propagators and Fokker-Planck equation).
Connection between the Langevin and  Fokker-Planck equations, as well as the interpretations of Langevin equation, %(and overdamped Langevin equation)
lead to various open questions, discussed in  \cite{Pavliotis,Kampen1981,Klafter1980, Sokolov2010}. 
%In  \cite{Kampen1981, Pavliotis} it was discussed how  Langevin equation corresponds to Fokker-Planck equation. 
As it was found in \cite{Serov}, when  diffusivity is not uniform,  overdamped Langevin equation's solution   depends on the interpretation of stochastic term that appears in it. % For physical systems in equilibrium, the dilemma is resolved when the long-time distribution of particle locations follow the Boltzmann distribution. For systems out of equilibrium the question remains open, 	and one needs to identify the active and diffusive components.
% In \cite{Postnikov} Brownian  non-Gaussian diffusion in heterogeneous systems was studied.
%In \cite{Petit2019} classes of random walks on temporal networks are discussed, such as node-centric vs. edge-centric walks and regarding the clock: passive or active walks. 

%\blue{What did we do in this paper?}
In section \ref{sec_cases} of our manuscript we considered the continuous HCTRW model starting with special cases of HCTRW in 1D. Calculations from Section \ref{sec_deriv} were made for the HCTRW without specification of underlying graph type. 
In particular, we studied the GME  for HCTRW on tree graphs (the most straightforward generalisation of a linear graph), on lattice graphs. If a graph is a regular lattice with  $\deg(i)=const$ $ \forall i$, then Eq.~(\ref{eq_GMEgen}) can be continuazed using methods described above. However for non-regular graphs this procedure should be done differently.
Another question is setting HCTRW model on a particular type of graph with temporal structure $\tau(x)$ so that the continuous version would satisfy  the diffusion equation
\begin{align}
\frac{d P_{x_0x}(t)}{d t} = \frac{d}{d x} \mathcal{D} \frac{dP_{x_0x}(t)}{dt} ,
\end{align}
where $\mathcal{D} =diag(D_1,...D_N)$ is the diffusion tensor, corresponding to the isotropic inhomogeneous  $N$-dimensional media with $D_i\neq D_j, i\neq j$. 

Moreover, in our manuscript we discuss the question about the relation between macroscopic view encoded in HCTRW interpretations and microscopic properties of the HCTRW model.
 First, we showed that the results for homogeneous cases of the HCTRW model, when all travel time distributions are the same,  correspond well to the results from \cite{Chechkin2006}. 
Then in Section  \ref{sec_cases} we derived diffusion equations, which generalize some previous findings on continuous limits for homogeneous random walk models \cite{Barkai} and heterogeneous random walk model \cite{GrebTupikina}. Moreover, this allows us to study the influence of microscopic heterogeneity, encoded through travel time distributions, on the macroscopic level of diffusion. 
The GME for homogeneous and heterogeneous cases of HCTRW, derived in the manuscript,  coincides well with the GME %Eq.~(\ref{eq_fin_GME}) 
found in \cite{Kenkre1973}. %, obtained here for general heterogeneous cases of HCTRW.
In subsection  \ref{sec_third} we considered less general case, when travel time distributions depend only on one site $\psi_{\bar{x}x'}=\psi_{x'}$ and derived kernel functions for exponential travel time distributions.
Comparing memory kernels of integro-differential operators  allows us to compare diffusion equations. 
%Furthermore, we incorporate additional parameters in order to dissect the interpretation of time and space dependent diffusion coefficients \cite{Sokolov2010}.  
%\blue{(insert explanation from [Serov et al. paper, abstract])}
In particular, we propose \emph{HCTRW continuous limits conjecture} about the relation between the HCTRW interpretations and \emph{various formalisms of diffusion equation}. %to link each of the interpretation of HCTRW model, described above,  with either Ito or Stratonovitch conventions, 
The \emph{HCTRW continuous limits conjecture} can be formulated as follows:
Ito formalism corresponds to the HCTRW model with travel time distributions  $\psi_{\bar{x}x'}$ depending only on nodes, where a random walk starts;
H\"anggi-Klimontovich - when $\psi_{\bar{x}x'}$ is evaluated only in nodes $x'$, where a random walk ends;  Stratonovich - when  $\psi_{\bar{x}x'}$  is evaluated in both nodes, or in function from both nodes, e.g. $ (\bar{x}+x')/2$. 
New travel time distributions can be set, for instance, 
as $\psi_{\bar{x}x'}(t) = \psi(\alpha x +(1-\alpha)x',t)$.
The first interpretation of HCTRW can be mapped to the CTRW model with $\psi_{\bar{x}}(t)$, where travel time distribution depends only on a starting point $\bar{x}$ and not on a node $x'$. 

\subsection{Langevin equation and SDE}

%references on filtration theorems 
%Our conjecture is related to the connection between HCTRW interpretations and diffusion equation formalisms, known as Ito, Stratonovich or H\"anggi formalisms. 
Another important class of models for studying diffusion formalisms are so-called barrier and accordion models \cite{Sokolov2010},
where often you can use  distribution of potential $U(x)$ and distribution of diffusion coefficients  $D(x)$. 
When $U(x)$ has local minimums of the same height,  we obtain a stationary distribution with a particle staying on the same height.
Hence for barrier and accordion models we get completely different behaviour (Stratonovich or H\"anggi), than for a trap model (Ito).
Another important argument about continuous limits of the HCTRW model is that initially travel time distributions
$\psi_{xx'}(t)$ do not depend on a form of matrix entries $Q_{xx'}$, although these entries can be independent, or be interrelated.
Transition matrix entries $Q_{xx'}$, in fact, can be related to the form of transition probability densities $\psi_{xx'}(t)$. %and rough paths theory 
In Langevin equation 
\begin{align}
\dot{x}(t) = f(x) + g(x,t)l(t)
%\label{eq_lang}
\end{align}
function $f(x)$ and $g(x)$ are two given functions, $l(t)$ is the rapid fluctuations. 
As it is pointed out in \cite{Kampen1981} the proper physical meaning should be given and we can distinguish the formalisms as follows. 
According to  stochastic differential equation (SDE) above, % $ \dot{x}(t) = f(x) + g(x,t)l(t)$, 
each pulse in $l(t)$ leads to a jump in $x$. That has an effect that a value $x$ to be used in $g(x,t)$ is undetermined (and hence also a size of a jump). 
The general form of Fokker-Planck equation (FP) can be written as: 
\begin{align}
\frac{d P_{x_0x}(s)}{dt} = 
\frac{d }{d x}\big[-f(x) P_{x_0x}(s) +
\\ \nonumber
\alpha (\frac{d}{d x}D(x)) P_{x_0x}(s) + D \frac{d P_{x_0x}(s)}{dx} \big], 
\end{align}
where different values of $\alpha$ correspond to: H\"anggi formalism for  $\alpha=0$, Stratonovich for $\alpha =1/2$, and Ito for $\alpha =1$. %Function  $f(x)$ is discussed below. 
As it was stressed in \cite{Sokolov2010}, the fact that $\alpha $ mathematically defines the position of a sampling point within the integration interval, 
is quite secondary and has essentially to do not with (non)-anticipation but with spacial symmetries of transition rates. % (\blue{add discussions on kinetic nature of parameter $\alpha$}). 
However for now the question stays, which formalism is more suitable for which random walk. 
In relation to this we  briefly discuss below the role of symmetries and random walk transition rates.
It is known that Ito integral is mathematically convenient 
to use  \cite{Gardiner} and that its martingale property allows to simplify derivation of Fokker-Planck equation from SDE.
However physically this choice is not always motivated, since in 
the physical system a term  $l(t)$ from Langevin equation  may not necessarily be the white noise \cite{Volpe} with finite correlation time. % hence $\langle l(t), l(t')\rangle =e^{|t-t'|}$.

\subsection{Discussions of the HCTRW interpretations}
The first interpretation of the HCTRW model was discussed in detail in Section \ref{sec_disc}, therefore we start with the second interpretation. 
The second interpretation of the HCTRW model with the exponential travel times  can be mapped  to the barrier model \cite{Thiel2016}. 
Then the arguments from \cite{Sokolov2010} for the barrier model help us to show that the second continuous HCTRW interpretation corresponds to Stratonovich formalism.
In the original barrier model \cite{Bouchard}
the transition rates depend on the barrier between nodes: $ \omega_0 q^{-\beta U_B(x_i,x')}$, where $U_B(x_i,x')$ is the barrier energy  between  nodes $x_i$ and $x'$. 
If the nodes would be exchanged, no changes in the equilibrium distribution $p_{eq}$ would take place. This brings the argument that a barrier model urges for H\"anggi-Klimontovich interpretation of the corresponding Langevin equation.
The main difference between the barrier and the trap model, illustrated on Fig.~\ref{fig_barri}, is that in the barrier model each node has zero potential. 
In Fig.~\ref{fig_barri} we illustrate the one dimensional barrier and trap models, 
which can be viewed as particular cases of HCTRW. 
%As pointed out in  \cite{Sokolov2010}  in the trap model  the potential $V_{max}$ in a combination $\exp (V_{max}/kT)$,  its second derivative $\omega_{max}$ enter $D$ but does not appear in equation\blue{(move to discussions)}
\\
The relation between the trap model \cite{Sokolov2010} and the HCTRW
should be explored in more details elsewhere. %Appendix
Moreover, it can be used for investigation of correspondence between conventions.

\section{Conclusions}
\label{sec_concl}

In this manuscript we presented the possible generalisation of the HCTRW model in continuous space and time.
Moreover, we describe in details possible relations between continuous and discrete quantities of RW, such as waiting time (or travel time) distributions $\Psi_{\bar{x}}(t)$  \cite{GrebTupikina}, and kernel functions
$\sum_{x'}Q_{\bar{x}x'}\psi_{\bar{x}x'}(t)=\mathcal{Q}_{\bar{x}}(t)$. This allows to open discussions on connections between discrete (in space) random walks with their continuous analogue of diffusion processes.
General analysis of continuous limits of random walk models on graphs deepens connections between kinetic properties and intrinsic quantities of diffusion equations. % therefore we are  open for future exploration of this topic.

\section{Outlook}
\label{sec_out}
As an outlook we plan to work on the derivation for a general case of second continuous HCTRW interpretation. %section III "stepping stone" 
In general, the HCTRW model and its continuous interpretations can be studied in various contexts, for instance on infinite  graphs with locally finite properties:  $\deg_{\bar{x}} < \infty$ $\forall \bar{x}$ of a graph. % which spectral properties are studied in   \cite{Mohar1989}. %(a finite chain with reflecting boundary conditions is a particular case)  % 
 Another point to consider is the case of HCTRW with various waiting time distributions, such as Sibuya distribution \cite{Angstmann2015}, which could lead to interesting specific properties of diffusion equation.
On another hand, following standard derivation of the Montroll-Weiss equation \cite{montroll1965} one can also study the non-markovian nature on the generalized master equation \cite{Hoffmann}.

An interesting special case to consider is the HCTRW model, where 
 	$Q_{xx'}$ transition matrix elements also depend on form of functional matrix entries 	$\psi_{xx'}(t)$, as it was also considered in a different context in the work on barrier models. %{Moutal2018}.
 	 One of the possible applications of the continuous limits of HCTRW to real-world systems includes 
 the models with permeable barriers %\cite{Moutal2018} 
 and the model with the space-dependent diffusion coefficient \cite{Lanoissele}. 
As for other possible applications
the fractional equations for heterogeneous HCTRW can be also applied to the fractional diffusion of ion channel gating
\cite{Goychuk2004}.
Studying spectral properties of infinite graphs \cite{Mohar1989}, on which HCTRW takes place, is another way to study continuous version of HCTRW.

The relation between \emph{various formalisms of diffusion equation and continuous interpretations of HCTRW} can 
be investigated further.  %to link each of the interpretation of HCTRW model, described above,  with either Ito or Stratonovitch conventions, 
In particular, Ito formalism in for the HCTRW  - when $\psi_{\bar{x}x'}$ depends only on $\bar{x}$;
Hanggi-Klimontovich - when $\psi_{\bar{x}x'}$ is evaluated only on $x'$;  Stratonovich - when  $\psi_{\bar{x}x'}$  is evaluated in between, i.e. in $ (\bar{x}+x')/2$. 
Other forms of travel time distribution can be set, for instance, 
	as $\psi_{\bar{x}x'}(t) = \psi(\alpha x +(1-\alpha)x',t)$.
	The first interpretation of HCTRW can be mapped to the CTRW with $\psi_{\bar{x}}(t)$, where the travel time depends only on the starting point $\bar{x}$ and not on $x'$.  Note, that such mapping between HCTRW given by set of $\psi_{\bar{x}}(t)$ and $\psi_{\bar{x}x'}(t)$ is not bijective. 
	In order to come from the general setup of HCTRW to the first interpretation one can also put $\psi_{\bar{x}}(t)=n_c\sum_{x'}\psi_{\bar{x}x'}(t)$, where $n_c$ is the normalisation constant.

\emph{Acknowledgements:} L.T. thanks Denis~S.~Grebenkov
from
Laboratoire de Physique de la Mati\`{e}re Condens\'{e}e (UMR 7643), CNRS -- Ecole Polytechnique, 91128 Palaiseau, France for inspiration and discussions of the main ideas of the manuscript. 
L.T. also acknowledges project \url{http://inadilic.fr/} and the support under Grant No. ANR-13-JSV5-0006-01 of the French National Research Agency. L.T. thanks support of Centre of Research and Interdisciplinarity in France and personally A.Serov, C.Vestergaard, G. Volpe and Y. Lanoissele for relevant suggestions of citations.

\end{document}